\documentclass[letterpaper, 10 pt, conference]{ieeeconf}  

\IEEEoverridecommandlockouts                              
\usepackage{graphicx}      
\usepackage{amsmath,amssymb}
\usepackage{setspace}
\usepackage{enumerate}
\usepackage{float}
\usepackage{dblfloatfix}    
\usepackage{comment}
\usepackage{curve2e}
\usepackage{multirow}
\usepackage{subfigure}
\usepackage{graphicx}
\usepackage{mathtools, cuted}
\usepackage{caption}
\usepackage{booktabs} 
\usepackage[italian,english]{babel}

\usepackage[T1]{fontenc}
\usepackage[utf8]{inputenc}
\usepackage{multirow}
\usepackage{mathrsfs}
\usepackage{framed}
\usepackage{url}
\newfam\bbfam
\font\tenbb=msbm10
\textfont\bbfam=\tenbb

\usepackage{algorithm}
\usepackage{algorithmic}

\usepackage{amssymb}	
\usepackage{cases}
\definecolor{rouge}{rgb}{1,0,0}
\definecolor{bleu}{rgb}{0,0,1}
\definecolor{vert}{rgb}{0,1,0}

\title{\LARGE \bf Software Rejuvenation for Secure Tracking Control} 

\author{$\text{Raffaele Romagnoli}^\dagger\text{, Bruce H. Krogh}^\ddagger\text{, Dionisio de Niz}^\ddagger\text{ and Bruno Sinopoli}^\dagger $\\
$^\dagger\text{Dept.of Electrical and Computer Engineering}$ \\
$^\ddagger\text{Software Engineering Institute}$\\
Carnegie Mellon University \\
Pittsburgh, PA USA \\
\{rromagno$|$dionisio$|$krogh$|$brunos\}@andrew.cmu.edu}

\begin{document}

\maketitle
\thispagestyle{empty}
\pagestyle{empty}

\begin{abstract}
Software rejuvenation protects cyber-physical systems (CSPs) against cyber attacks on the run-time code by periodically refreshing the system with an uncorrupted software image. The system is vulnerable to attacks when it is communicating with other agents.  Security is guaranteed during the software refresh and re-initialization by turning off all communication.  Although the effectiveness of software rejuvenation has been demonstrated for some simple systems, many problems need to be addressed to make it viable for real applications. This paper expands the scope of CPS applications for which software rejuvenation can be implemented by introducing architectural and algorithmic features to support trajectory tracking. Following each software refresh, while communication is still off, a safety controller is executed to assure the system state is within a sufficiently small neighborhood of the current point on the reference trajectory. Communication is then re-established and the reference trajectory tracking control is resumed. A protected, verified hypervisor manages the software rejuvenation sequence and delivers trusted reference trajectory points, which may be received from untrusted communication, but are verified using an authentication process. We present the approach to designing the tracking and safety controllers and timing parameters and demonstrate the secure tracking control for a 6 DOF quadrotor using the PX4 jMAVSim quadrotor simulator.  The concluding section discusses directions for further research.   
\end{abstract}

\section{INTRODUCTION}
Software rejuvenation, an established method to improve the robustness against software faults in traditional computing systems \cite{ABL+12}, has been proposed recently to deal with unmodeled and undetectable cyber attacks on cyber-physical systems (CPSs) \cite{ASW17,arroyo2017fired}.   The idea is to periodically refresh the run-time system completely with a trusted, secure copy of the control software to thwart attacks that may have changed the on-line code.  Although the basic concept has been implemented for some demonstration systems, applications have been limited to cases where all of the information required for control is available a priori so that it can be reloaded with each software refresh. This paper presents an architecture and algorithms to support secure tracking control using software rejuvenation when the reference trajectory points are being updated from possibly insecure communication throughout the execution of the mission.

The concept of software rejuvenation was introduced by Huang et al. in 1995 \cite{HKKF95} to address the problem of so-called software aging; that is, failures that occur when a running program encounters a state that was not anticipated when the software was designed.  The basic idea is to restart the software intermittently at a "clean" state, either through complete system reboot or by returning to a recent checkpoint, with the hope that this will prolong the time until unanticipated states occur that might cause failures. Since the introduction of the concept, there has been considerable research into the development and performance of software rejuvenation strategies \cite{CNPR14}, and it has become a practical tool for enhancing the robustness of many computing systems \cite{ABL+12}.

Although software aging remains the primary motivation for implementing software rejuvenation strategies in computing systems, a few papers have proposed software rejuvenation to enhance system security \cite{AP04,LKGK14}. In contrast to software aging where mean-time to failure can be the basis for timing software refresh, the frequency of software refresh to defend against malicious attacks must be determined by the length of time a system can remain viable once its security has been violated. 

Arroyo et al. \cite{ASW17,arroyo2017fired} propose software rejuvenation as a strategy for CPS security and demonstrated the concept for a quadrotor controller and an automotive engine controller. These examples illustrate how refreshing software in a CPS impacts performance and introduces timing  constraints and safety considerations that are not present in traditional computing systems.  

Abidi et al. \cite{Abdi2018} develop software rejuvenation for CPS further by introducing three concepts.  First, the \textit{hardware root of trust} is a secure onboard module that hosts the capabilities that must be available without compromise to implement software rejuvenation. Second, the \textit{secure execution interval} (SEI) is a period during which all external communication is disabled so that no cyber attacks can occur as the software is refreshed.  Third, the \textit{safety controller}, which executes immediately following a software refresh and during the SEI, drives the CPS to a known safe state before restoring communication and returning the system to mission control with vulnerability to attacks.  Abidi et al. use a simple, conservative reachability algorithm to determine the time that can be allowed before the next software reset and illustrate their approach for a simulated warehouse temperature controller and a bench-top 3-DOF helicopter.

In this paper, we adopt the overall approach to software rejuvenation from \cite{Abdi2018}, but modify the timing strategy based on analysis of the control problem. We also use the concept of a secure, trusted hypervisor introduced by Vasudevand et al. \cite{VCM+16}, rather than separate hardware, to provide the required secure computing components.  We formulate the software rejuvenation problem for tracking control applications, which arise in many CPS applications.  Using invariant sets to design the safety controller, an event-based algorithm to switch reference points for the tracking controller, and data authentication to manage reference points received from possibly compromised communication, we demonstrate how software rejuvenation can be used to achieve provable operational safety and robustness against cyber attacks that take over the system control by changing the run-time software. We illustrate our contributions using simulation of the nonlinear dynamics for a 6-DOF quadrotor.

The paper is organized as follows. Section \ref{sec:TrackingControl} describes the standard architecture of hierarchical tracking control systems and the event-based method for switching points to be tracked along a reference trajectory. Section \ref{sec:SafteyControl} presents the design of safety controllers and timing for software rejuvenation based on invariant sets.  Tracking control and safety control are combined in Sec. \ref{sec:SoftwareRejuvenation} to implement the software rejuvenation algorithm using the hypervisor architecture. The concepts are applied in Sec. \ref{sec:example} to implement a secure tracking control system for a 6 DOF drone and simulation results illustrate the tracking control performance when the system is subjected to a cyber attack. An appendix provides the mathematical details for the drone control application. The concluding section discusses directions for further research and development. 

\section{TRACKING CONTROL}\label{sec:TrackingControl}
\begin{figure}[h!]
 \includegraphics[scale=.60,trim=45 120 160 100,clip]{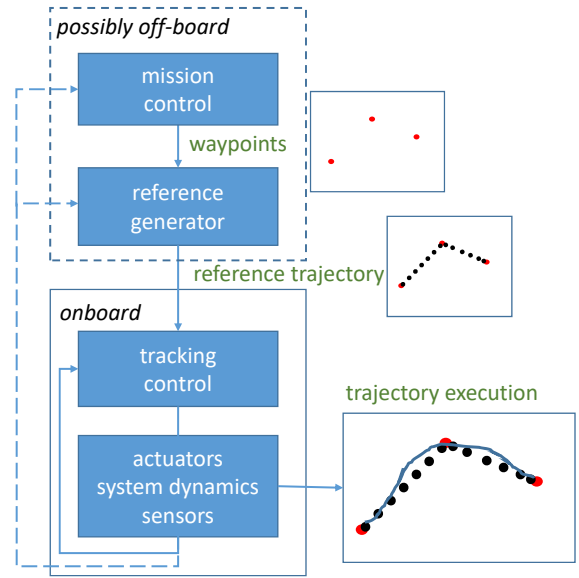}   
\caption{Hierarchical tracking control systems.}\label{fig:TrackingControl}
\end{figure}

CPSs that perform missions involving motion are typically operated using a hierarchical control architecture, as illustrated in Fig. \ref{fig:TrackingControl}. A \textit{mission controller} defines a sequence of \textit{waypoints} to be visited.  A \textit{reference generator} produces a sequence of points called the \textit{reference trajectory} to take the CPS from waypoint to waypoint along a safe and feasible path.  Motion along the reference trajectory is executed by a \textit{tracking controller} that generates the commands to the system actuators.  Depending on the application, the mission controller and reference governor may be onboard or off-board, but the trajectory controller usually needs to be onboard to satisfy the real-time constraints for executing the trajectory.  This paper focuses on the implementation of the tracking controller.

The tracking controller is typically implemented as a regulator that will drive the system to a given set point, and motion is accomplished by switching along the points on the reference trajectory.  Since it is not necessary to actually stop at the points on the reference trajectory, the switching from the current reference point to the next reference point can be implemented using a time-based or event-based scheme. In time-based schemes, the points on the reference trajectory have time stamps and the reference point for the tracking controller is updated as time evolves.  In event-based schemes, the reference point for the tracking controller is updated when the system is within a given neighborhood of the current reference point.  The secure tracking control proposed in this paper uses the following event-based scheme to update the reference points for the trajectory controller.

\begin{figure}[!h]
\begin{center}
 \includegraphics[scale=.50,trim=90 180 60 180,clip]{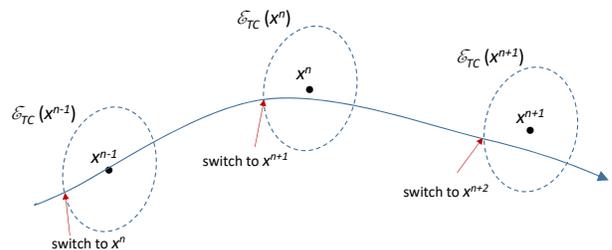}   
\end{center}
\caption{Event-based reference point update for tracking control.}\label{fig:RefUpdate}
\end{figure}

We assume the reference points along the reference trajectory are equilibrium states for the system.  Let $x^{n}$ denote the current equilibrium reference point and suppose that a feedback control law is implemented to drive the state of the system to $x^n$ and stabilize it at that equilibrium set. If $x^{n}$ is not the final point in the reference trajectory (the final waypoint), the reference point can be updated to the next reference point, denoted by $x^{n+1}$ before $x^{n}$ is reached.  The condition for switching the reference point to $x^{n+1}$ is defined by a region about the current reference point, denoted by $\mathcal{E}_{TC}(x^{n})$. Figure \ref{fig:RefUpdate} illustrates this event-based rule for updating the reference point for tracking control.  

\section{SAFETY CONTROL}\label{sec:SafteyControl}
This section describes the safety controller that plays a key role in achieving secure tracking control using software rejuvenation. The safety controller is designed to guarantee the state of the system remains within a safe neighborhood of the current equilibrium point being used as the reference point by the tracking controller, even if there is a cyber attack. Let $x_{eq}$ denote the current equilibrium point for tracking control, which we will refer to as \textit{normal control} in this section. The safe neighborhood of $x_{eq}$, denoted by $\mathcal{E}_C(x_{eq})$, is a set of states satisfying the operating constraints as well as the range of states around $x_{eq}$ for which the safety controller is guaranteed to work properly.   

During normal control, the system is vulnerable to cyber attacks, which means that the run-time code may be modified and the control actions can drive the system away from $x_{eq}$. The controller is also disabled during a software refresh, which is performed before safety control can be executed.  We call this period of normal control followed by the software refresh the period of \textit{uncertain control}, with the duration denoted by $T_{UC}$.

To assure the state of the system does not exit $\mathcal{E}_C(x_{eq})$ during the period $T_{UC}$, it must be guaranteed the state is in a sufficiently small neighborhood of $x_{eq}$ before normal control (and vulnerability to attacks) is initiated such that the state trajectory cannot be driven out of $\mathcal{E}_C(x_{eq})$ with \textit{any} control action available to an attacker. We let $\mathcal{E}_SC(x_{eq})$ denote this smaller neighborhood of $x_{eq}$, and note that $\mathcal{E}_C(x_{eq})$ must satisfy one additional condition. That is, if safety control is initiated when the system is at any state in $\mathcal{E}_C(x_{eq})$, it must be the case that under safety control, the system is taken to $\mathcal{E}_{SC}(x_{eq})$ in finite time along a trajectory that does not leave $\mathcal{E}_C(x_{eq})$. Moreover, the safety controller must keep the state trajectory in $\mathcal{E}_SC(x_{eq})$ once it has entered $\mathcal{E}_SC(x_{eq})$.  In control terminology, it is necessary for $\mathcal{E}_C(x_{eq})$ and  $\mathcal{E}_SC(x_{eq})$ to be \textit{invariant} under safety control.

The requirements for $\mathcal{E}_C(x_{eq})$ and $\mathcal{E}_SC(x_{eq}) \subset \mathcal{E}_C(x_{eq})$ are summarized by the following technical conditions:
\begin{enumerate}[i]
    \item $\mathcal{R}(\mathcal{E}_C(x_{eq}),t;SC) \subseteq \mathcal{E}_C(x_{eq}) \  \forall \ t > 0$, where $\mathcal{R}(\mathcal{E}_C(x_{eq}),t;SC)$ is the set of states reached at time $t$ applying safety control (SC) starting at time $t=0$ from any state in $\mathcal{E}_C(x_{eq})$;\footnote{We assume the system dynamics are time-invariant.}
    \item $\exists \ \bar{T}_{SC}>0 \  \ni \  \mathcal{R}(\mathcal{E}_C(x_{eq}),t;SC) \subseteq \mathcal{E}_{SC}(x_{eq}) \  \forall \ t \geq \bar{T}_{SC}$; and
    \item $\exists \  T_{UC} > 0 \ \ni \  \mathcal{R}(\mathcal{E}_SC(x_{eq}),t;\mathcal{U}) \subseteq \mathcal{E}_C(x_{eq}) \ \forall \ t < T_{UC}$.
\end{enumerate}

\begin{figure}[!h]
\begin{center}
 \includegraphics[scale=.55,trim=180 180 60 185,clip]{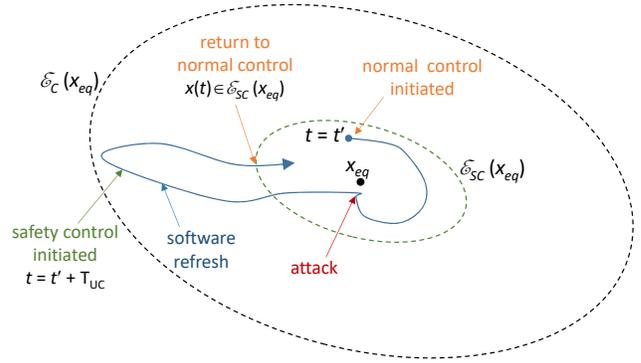}   
\end{center}
\caption{Safety control: initiated after the period $T_{UC}$ since the last initiation of normal control; executed until the state is in $\mathcal{E}_{SC}(x_{eq})$, before returning the system to normal control.}\label{fig:SafetyControl}
\end{figure}

Figure \ref{fig:SafetyControl} illustrates the definitions and conditions described above. Section \ref{sec:example} presents methods for finding $\mathcal{E}_C(x_{eq})$,  $\mathcal{E}_{SC}(x_{eq})$ and $T_{UC}$ for a drone control application.

\section{SOFTWARE REJUVENATION FOR TRACKING CONTROL}\label{sec:SoftwareRejuvenation}
This section describes the architecture and the algorithm to achieve secure tracking control using software rejuvenation using the tracking controller and safety controller described in the previous sections. 

\begin{figure*}[!ht]
\begin{center}
 \includegraphics[scale=.50,trim=35 140 50 150,clip]{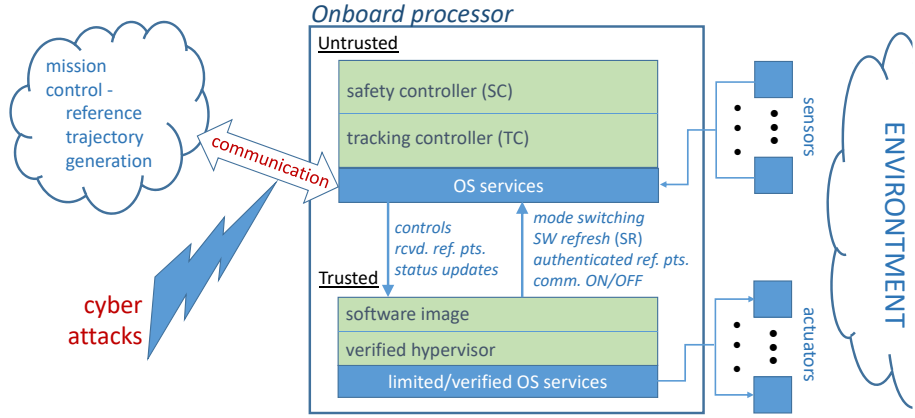}   
\end{center}
\caption{System architecture to support secure software rejuvenation for tracking control.}\label{fig:Processor}
\end{figure*}

Figure \ref{fig:Processor} presents the system architecture.  The onboard processor has two partitions, trusted  and  untrusted.  The \textit{trusted partition} has protected memory and services that cannot be accessed by any processes in the untrusted partition.  It hosts the software image that will refresh the run-time code and memory in the untrusted partition.  It also hosts a verified, trusted hypervisor \cite{VCM+16}, which manages the software rejuvenation cycle, initiates and terminates processes in the untrusted partition, authenticates reference points for the tracking controller, and processes control commands from the controllers to be delivered to the actuators (explained below).  The hypervisor uses a limited, verified set of OS services that are also hosted in the trusted partition.

The \textit{untrusted partition} hosts the safety and tracking controllers, the operating system and any other functions required to operate the system. This partition is untrusted because it is too complex to be verified formally and it receives communication from other agents, which includes sensors and off-board agents.  It is this communication, which is necessary for the system to execute its missions, which makes the system vulnerable to cyber attacks. As shown in Fig. \ref{fig:Processor}, although reference points may be received from the communication, they are not used directly by the controllers. The received reference points are passed to the hypervisor for authentication and the hypervisor manages the reference point update, as described below. 

\begin{figure*}[!ht]
\begin{center}
 \includegraphics[scale=.60,trim=55 180 50 150,clip]{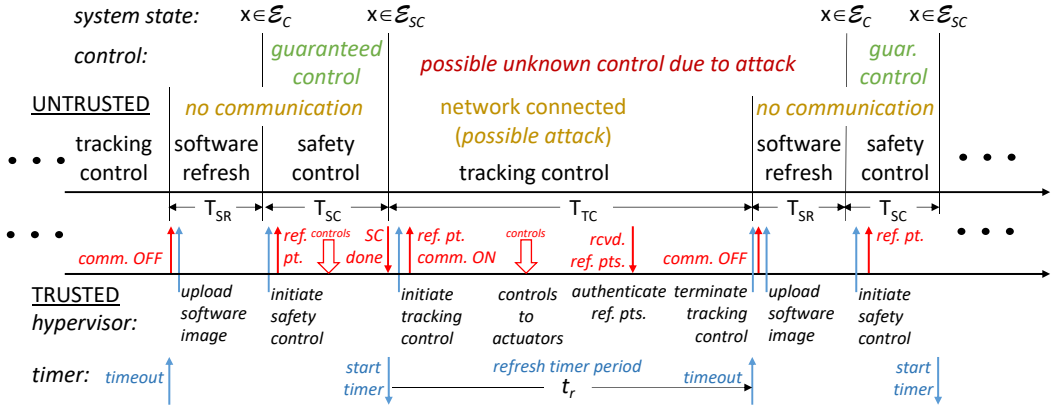}   
\end{center}
\caption{Software rejuvenation control modes.}\label{fig:TimingDiagram}
\end{figure*}

Figure \ref{fig:TimingDiagram} details the timing for the software rejuvenation cycle and the communication between the trusted and untrusted partitions.  Before a software refresh, the hypervisor turns off the communication to the untrusted partition, making it invulnerable to cyber attacks.  The communication remains off through the execution of the safety control, which follows immediately after the software refresh. As indicated in the figure, the safety control executes until the state has been driven into  $\mathcal{E}_{SC}$ for the current reference point. A timer is set and tracking control is then initiated by the hypervisor, which provides the reference point to the tracking controller. Any new reference points received during tracking control are passed to the hypervisor for authentication.  Tracking control is terminated either when the state reaches $\mathcal{E}_{TC}$ for the current reference point or the timer times out after the timer period $t_r$ (only the timeout case is shown in Fig. \ref{fig:TimingDiagram}). Communication is then turned off and software refresh is executed again. The software refresh takes a known amount of time, $T_{SR}$, so by choosing $t_r \leq T_{UC}-T_{SR}$, it is guaranteed the state is in $\mathcal{E}_C$ when the safety control is initiated, as indicated in the figure.

The hypervisor uses algorithm \ref{alg:RefPoints} to update the reference points for the controllers.\footnote{Algorithm \ref{alg:RefPoints} shows only the functions of the hypervisor relevant to reference point management. }

\begin{algorithm}
\caption{Reference Point Update}\label{alg:RefPoints}
\begin{algorithmic}[1]
\STATE $x_{start}$ : given initial reference point
\STATE Initialize:
\STATE \hspace{0.2cm} COMMUNICATION OFF;
\STATE \hspace{0.2cm} $x_{cur} := x_{start}$ : current reference point
\STATE \hspace{0.2cm} SOFTWARE REFRESH
\STATE \hspace{0.2cm} Initiate Safety Control
\STATE \hspace{0.2cm} Send $x_{cur}$
\STATE \hspace{0.2cm} !Receive Safety Control Completed
\STATE REPEAT 
\STATE \hspace{0.2cm} COMMUNICATION ON
\STATE \hspace{0.2cm} START TIMER with time $t_r$
\STATE \hspace{0.2cm} Initiate Tracking Control
\STATE \hspace{0.2cm} Send $x_{cur}$
\STATE \hspace{0.2cm} FINISHED := FALSE
\STATE \hspace{0.2cm} WHILE NOT FINISHED
\STATE \hspace{0.2cm} \hspace{0.2cm}IF Receive new reference point $x_{new}$
\STATE \hspace{0.2cm} \hspace{0.2cm} \hspace{0.2cm} IF NOT AUTHENTICATE($x_{new})$
\STATE \hspace{0.2cm} \hspace{0.2cm} \hspace{0.2cm} \hspace{0.2cm} $x_{new} := x_{cur}$
\STATE \hspace{0.2cm} \hspace{0.2cm} \hspace{0.2cm} ENDIF
\STATE \hspace{0.2cm} \hspace{0.2cm} IF Receive Tracking Control Completed
\STATE \hspace{0.3cm} \hspace{0.2cm} OR Receive TIMEOUT
\STATE \hspace{0.2cm} \hspace{0.2cm} \hspace{0.2cm} FINISHED := TRUE
\STATE \hspace{0.2cm} \hspace{0.2cm} ENDIF
\STATE \hspace{0.2cm} ENDWHILE
\STATE \hspace{0.2cm} COMMUNICATION OFF;
\STATE \hspace{0.2cm} SOFTWARE REFRESH
\STATE \hspace{0.2cm} Initiate Safety Control
\STATE \hspace{0.2cm} Send $x_{cur}$
\STATE \hspace{0.2cm} !Receive Safety Control Completed
\STATE \hspace{0.2cm} $x_{cur} := x_{new}$
\STATE END REPEAT
\end{algorithmic}
\end{algorithm}

\begin{figure*}[!h]
\begin{center}
 \includegraphics[scale=.6,trim=35 250 50 150,clip]{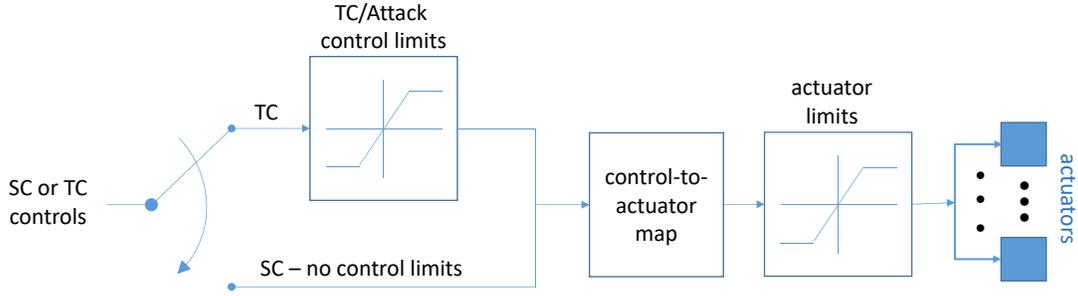}   
\end{center}
\caption{Hypervisor control signal processing.}\label{fig:ControlLimits}
\end{figure*}

As is illustrated in the following section, it is useful to constrain the control actions that can be taken during tracking control so that any attacks during this period are subjected to constraints that limit the rate at which an attacker might drive the system from $\mathcal{E}_{SC}$ to the boundary of $\mathcal{E}_{C}$. Enforcing such limits increases $T_{UC}$. These limits can be imposed by passing all controls through the hypervisor. On the other hand, the full control capability is available to the safety controller, which helps to minimize any time required for safety control actions. Figure \ref{fig:ControlLimits} illustrates this processing performed by the hypervisor to communicate the controls from the controllers to the actuators. Note that the hypervisor is a relatively small process that is fully verified, so it is only able to impose limits on the control signals.

\section{APPLICATION: DRONE CONTROL}\label{sec:example}
This section shows the application of the proposed software rejuvenation for secure tracking control for a 6-DOF Quadrotor  (or Unmanned Aerial Vehicle (UAV)) system. Specificall,y a Generic 10" Quadrotor + geometry airframe \cite{px4} with full thrust of 4 N, and full torque of 0.05 Nm for each motor is considered. The simulation are performed with the jMAVSim flight simulator.
A 6-DOF quadrotor is a nonlinear system described by a 12-dimensional state-space model that contains position, orientation and linear and angular velocities of the UAV. \cite{beard2008quadrotor}. There are four controls: the normal force to the airframe and three torques related to the orientation angles of the UAV. The bounds on the controls are:   the max total force normal to the air frame is $16$N, the max torque related to the roll and pitch angles is $\pm 0.66$ Nm, and $\pm 0.1$ Nm for the yaw.
Those controls are mapped by a matrix multiplication ("mixer") to the actuator signals for the four motors. Each signal is normalized with respect to the maximal power that can be generated by each motor. 

The UAV is stabilized by an LQR controller designed considering the linearized model of the system around an equilibrium point $x_{eq}$, that without loss of generality can be considered the origin of the state space \cite{araar2014full},\cite{bolandi2013attitude},\cite{bouabdallah2004pid}. See the appendix for the details of the computations discussed in the following subsections.


\subsection{Safety Controller}
The safety controller is the LQR controller described in the Appendix, using $Q=50\cdot I_{12}$ and $R=100 \cdot I_{12}$ where $I_{12}$ is the identity matrix. The safe set $\mathcal{C}$ is a convex polyhedral subset of the state space $\mathbb{R}^{12}$ which imposes the following constraints on:
\begin{itemize}
    \item the position: $\pm 2.5$ m along $z$ (vertical direction), and $\pm 1$ m along $x$ and $y$ (horizontal direction);
    \item the orientation: the roll $\phi$, pitch $\theta$ and yaw $\psi$ angles are bounded between $\pm \pi/4$;
    \item the linear velocities: $\pm 5$ m/s for the vertical and $\pm 2$ m/s for the horizontal directions;
    \item the angular velocities: $\pm 5$ rad/s for all the angles.
\end{itemize}

Writing the above set of constraints as in \eqref{C1}, and solving the maximization problem \eqref{safeinv}, the safe set $\mathcal{E}_{\mathcal{C}}(x_{eq})$ is the maximum volume ellipsoid contained in \eqref{C1} with the symmetric matrix $P>0$.

The computation of the overapproximation of the reachable set $\mathcal{R}^+(\mathcal{C}_{SC}(x_{eq}),T_{UC},\mathcal{U})$ \eqref{V+1}, for the verification of the admissible values of $\epsilon_{SC}$ and $T_{UC}$ requires $\mathcal{U}$ which is, in this case, a polytopic set generated by:
\begin{itemize}
    \item[1)] the bound on the torques related to the pitch and roll angles is $\pm 0.0033$ Nm, for yaw is  $\pm 0.0005$ Nm.
    \item[2)] the bound on the max total force is $14$ N, and the lowest admissible value in case of a sudden shut down of the propellers is $2$ N.
\end{itemize}

\begin{figure*}[!h]
	\centering
	\begin{subfigure}[]{} 
		\includegraphics[height=8cm,width=8cm]{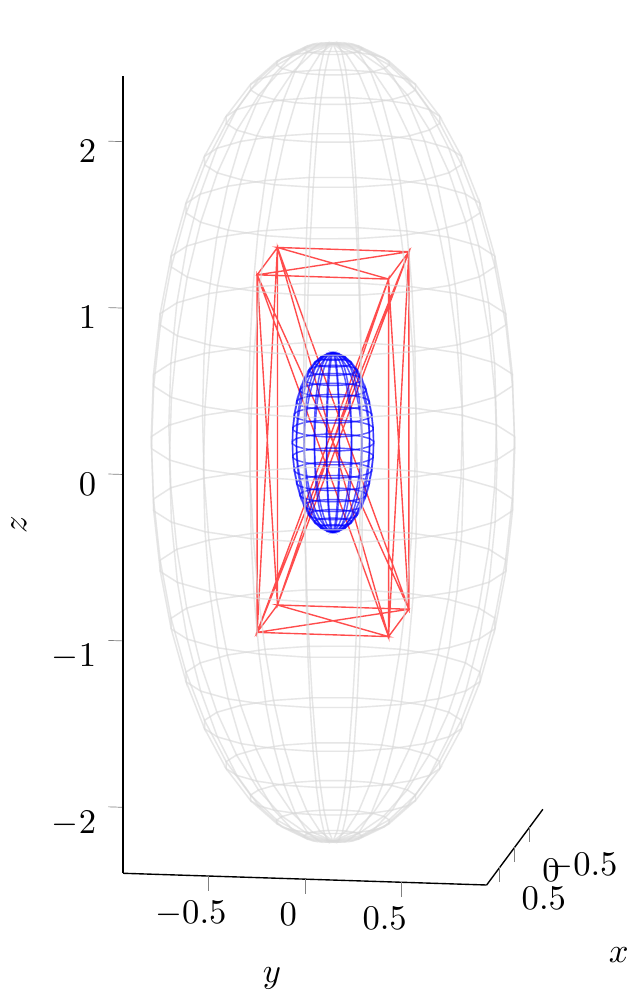}
	\end{subfigure}
	\begin{subfigure}[]{} 
		\includegraphics[height=8cm,width=8cm]{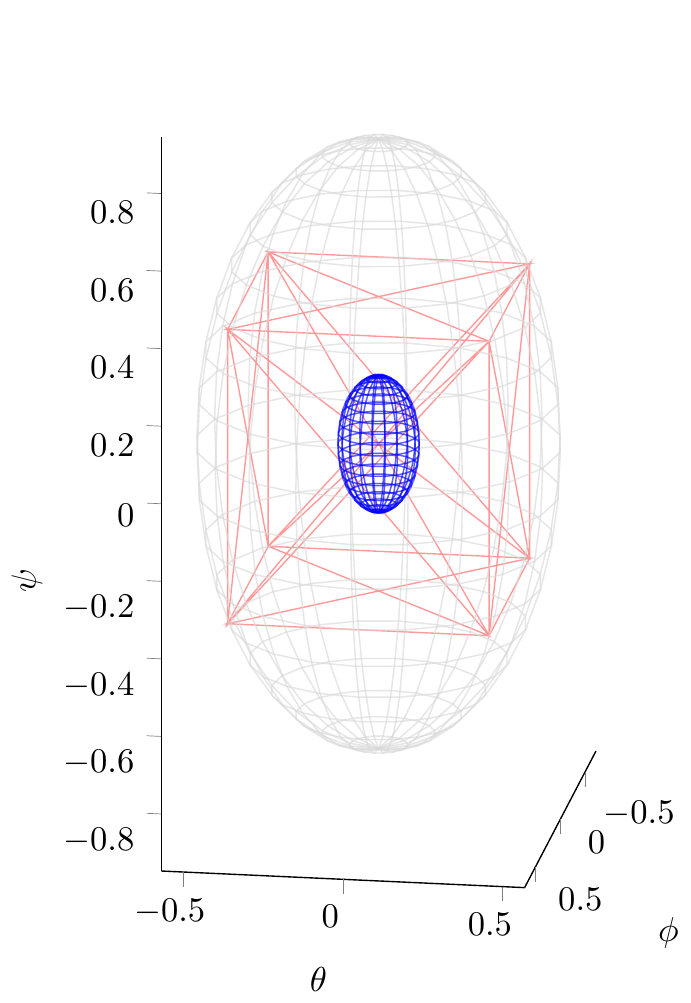}
	\end{subfigure}
	\caption{Projections of $\mathcal{E}_{\mathcal{C}}$(gray), $\mathcal{E}_{SC}$ (blue) and $\mathcal{R}^+(\mathcal{C}_{SC},T_{UC},\mathcal{U})$ computed for $\epsilon=0.05$ and $T_{UC}=0.18s$ (red line) on the position (left), and orientation (right).} 
	\label{fig:pos_inv_reach_set}
\end{figure*}

\begin{figure*}[!h]
	\centering
	\begin{subfigure}[]{} 
		\includegraphics[height=8cm,width=8cm]{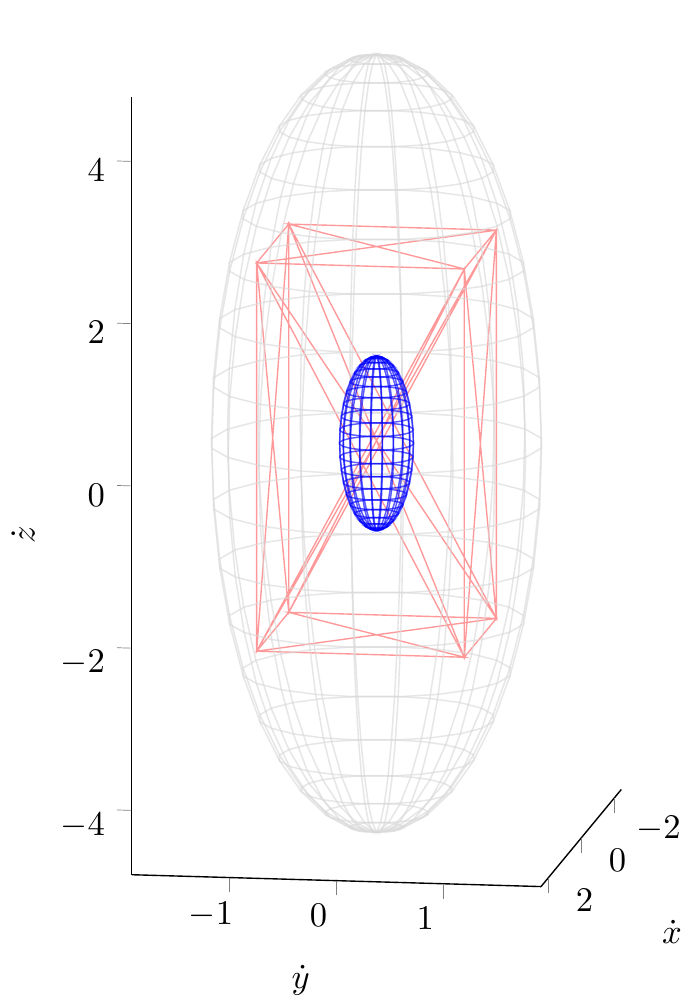}
	\end{subfigure}
	\begin{subfigure}[]{} 
		\includegraphics[height=8cm,width=8cm]{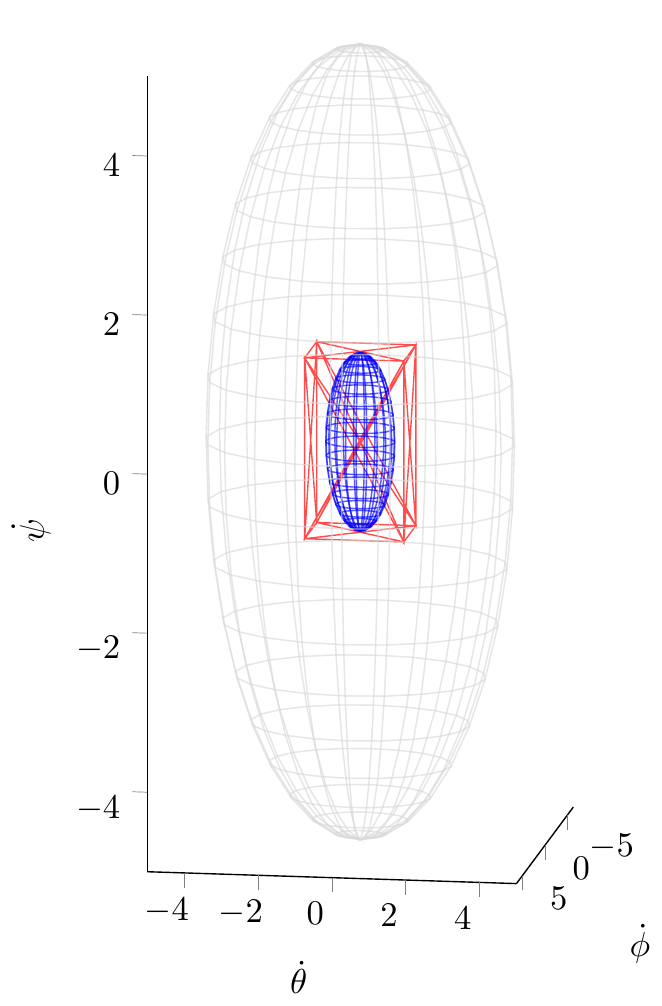}
	\end{subfigure}
\caption{Projections of $\mathcal{E}_{\mathcal{C}}$(gray), $\mathcal{E}_{SC}$ (blue) and $\mathcal{R}^+(\mathcal{C}_{SC},T_{UC},\mathcal{U})$ computed for $\epsilon=0.05$ and $T_{UC}=0.18s$ (red line) on the linear velocity space.}
\label{fig:vel_inv_reach_set}
\end{figure*}

Figures \ref{fig:pos_inv_reach_set} and \ref{fig:vel_inv_reach_set} show the sets $\mathcal{E}_{\mathcal{C}}$, $\mathcal{E}_{SC}$ and $\mathcal{R}^+(\mathcal{C}_{SC}(x_{eq}),T_{UC},\mathcal{U})$ projected in the space of the position, orientation, linear and angular velocities for $\epsilon_{SC}=0.05$ and $T_{UC}=0.18s$. In all four projections $\mathcal{R}^+(\mathcal{C}_{SC}(x_{eq}),T_{UC},\mathcal{U}) \subseteq \mathcal{E}_{\mathcal{C}}(x_{eq})$. In fact, the safe set $\mathcal{E}_{\mathcal{C}}(x_{eq})$ contains all the vertexes of  $\mathcal{R}^+(\mathcal{C}_{SC}(x_{eq}),T_{UC},\mathcal{U})$. It is important to note that, despite the conservative effects due to the bounds on the torques on the reachable set of the angular velocities (Fig. \ref{fig:vel_inv_reach_set}- right), linear velocities (Fig. \ref{fig:vel_inv_reach_set}- left) and angles (Fig. \ref{fig:pos_inv_reach_set}- right) have the vertexes of their reachable sets close to the border of $\mathcal{E}_{\mathcal{C}}(x_{eq})$.

\subsection{Tracking Control}

The goal of the tracking control is to follow a sequence of equilibrium points $x_{eq}$ to drive the system from a specified initial point to a specified final point (the two waypoints).
For the quadrotor, an equilibrium point corresponds to  hovering at a certain altitude, with zero roll pitch angles, and a given value of yaw. All the velocities are zero. A constant feedforward input is added to the controls in order to provide a sufficient normal force that compensates the effect of the g-force. In this case that value is $7.848$ N.

The mechanism generating the sequence of equilibrium points by the tracking control scheme depends on $\mathcal{E}_{TC}(x_{eq}) \subset \mathcal{E}_{SC}(x_{eq})$. When the system moves into $\mathcal{E}_{TC}(x_{eq})$, the tracking controller switches from $x_{eq}$ to the next one $x_{eq}'$ after the next reboot.
This example considers $\mathcal{E}_{TC}(x_{eq})$ as an ellipsoids described by the matrix $P>0$ with $\epsilon_{TC}=0.01$ and micro-reboots of $0.03$ s.
The mission of this simulation consists in moving the UAV from the position $(x=1,y=0,z=2,\psi=0)$ to $(x=1,y=2,z=4,\psi=0)$. 

\subsection{Simulation}

The aim of the simulation is to illustrate the effectiveness and robustness against attack of the proposed approach. To illustrate that, an attack that suddenly turns off the propellers is considered.

\begin{figure*}[!h]
\centering
\includegraphics[height=8cm, width=16cm]{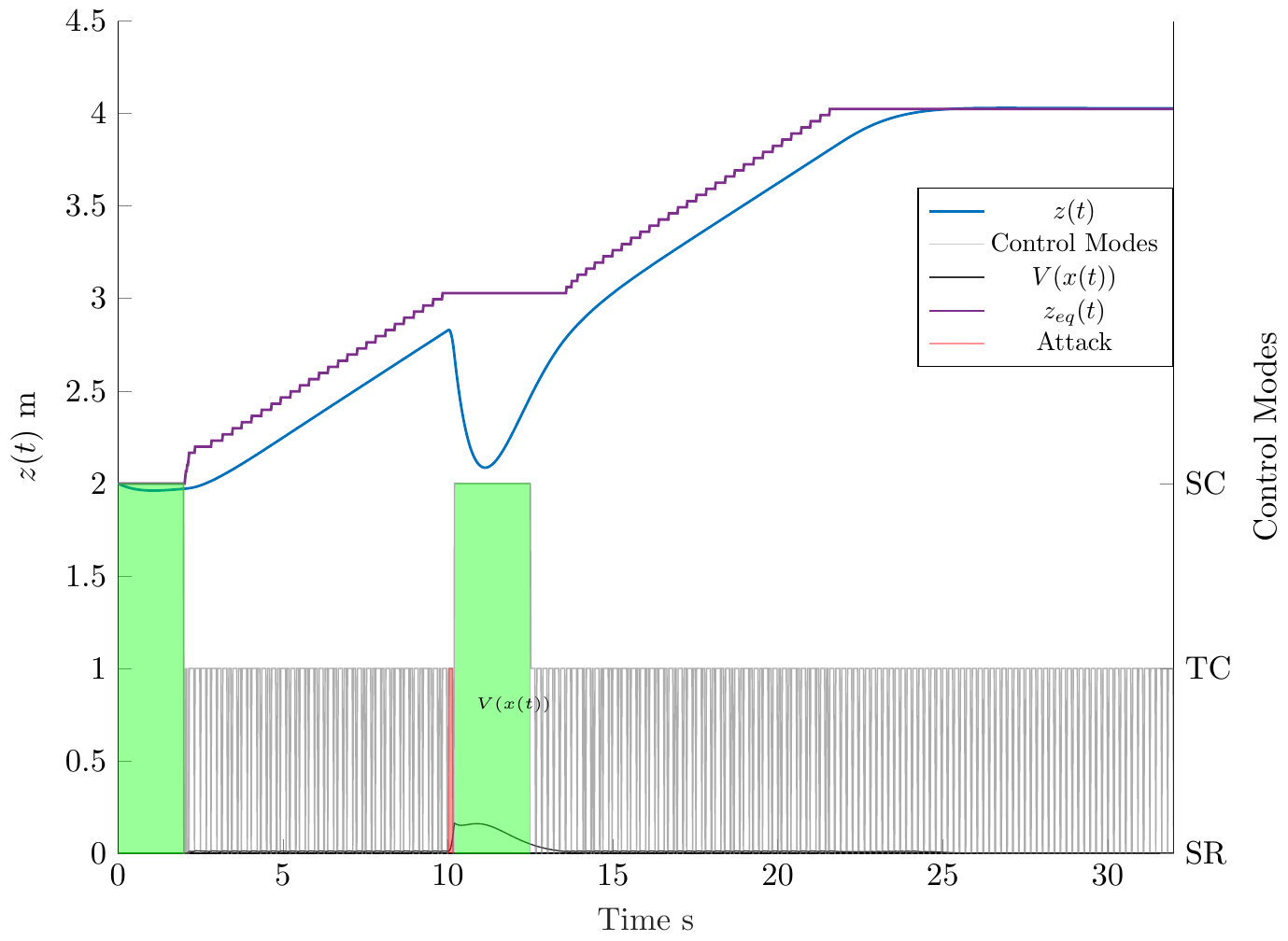} 
\caption{System response along $z$ respect to the time (blue), the equilibrium points (magenta), the switching between the several control modes (gray) and the Lyapunov function $V(x(t))$ (black).}
\label{fig:timeline}
\end{figure*}

Figure \ref{fig:timeline} describes the overall simulation considering projecting the position along $z$. Besides the actual behaviour of the UAV along $z$, it shows the equilibrium points updates, the switches between the several control modes (TC,SR,SC), and the value of the Lyapunov function $V$, respect to the time.
Before that the mission starts, the secure control is forced to run for 2s. After that, the mission control start to update the equilibrium points (magenta) if the the state of the system is in $\mathcal{E}_{TC}(x_{eq})$. The equilibrium points pull the altitude of the UAV towards the final equilibrium point. The current equilibrium point $x_{eq}$  switches to the next one $x_{eq}'$ before that the quadrotor gets to  $x_{eq}$. The gray plot represents the switches between micro-reboots and TC where the equilibrium points are updated. The attack happens during the period highlighted in red. Since the system is far away from the current equilibrium point, the reference point for the tracking controller does not update it until  the system reaches $\mathcal{E}_{TC}(x_{eq})$. The figure also shows the values of the Lyapunov function that are very small except when the attack occurs. 

\begin{figure*}[!h]
	\centering
	\begin{subfigure}[]{} 
		\includegraphics[height=8cm, width=8cm]{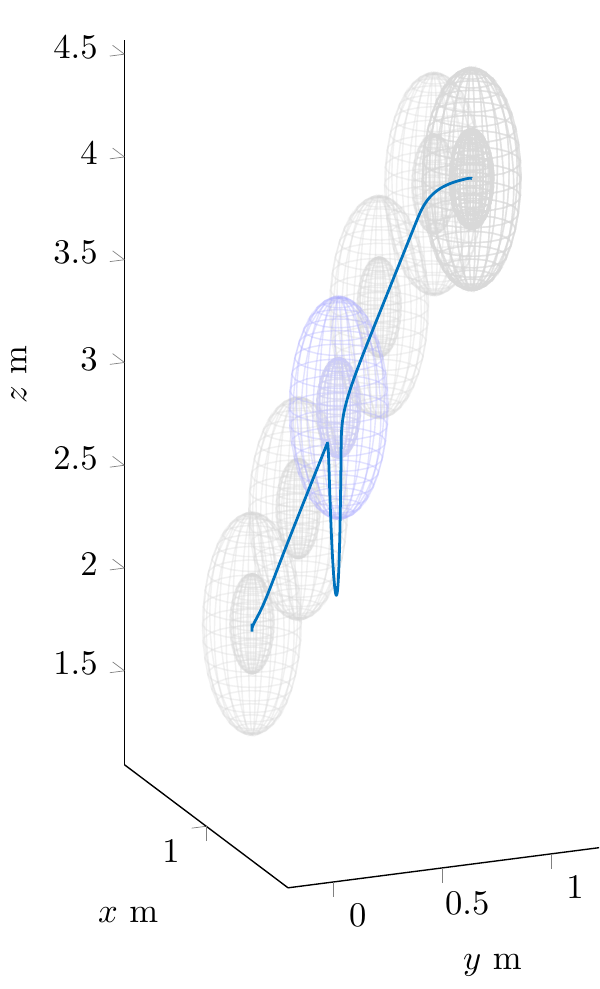}
	\end{subfigure}
	\hspace{2em} 
	\begin{subfigure}[]{} 
		\includegraphics[height=8cm, width=8cm]{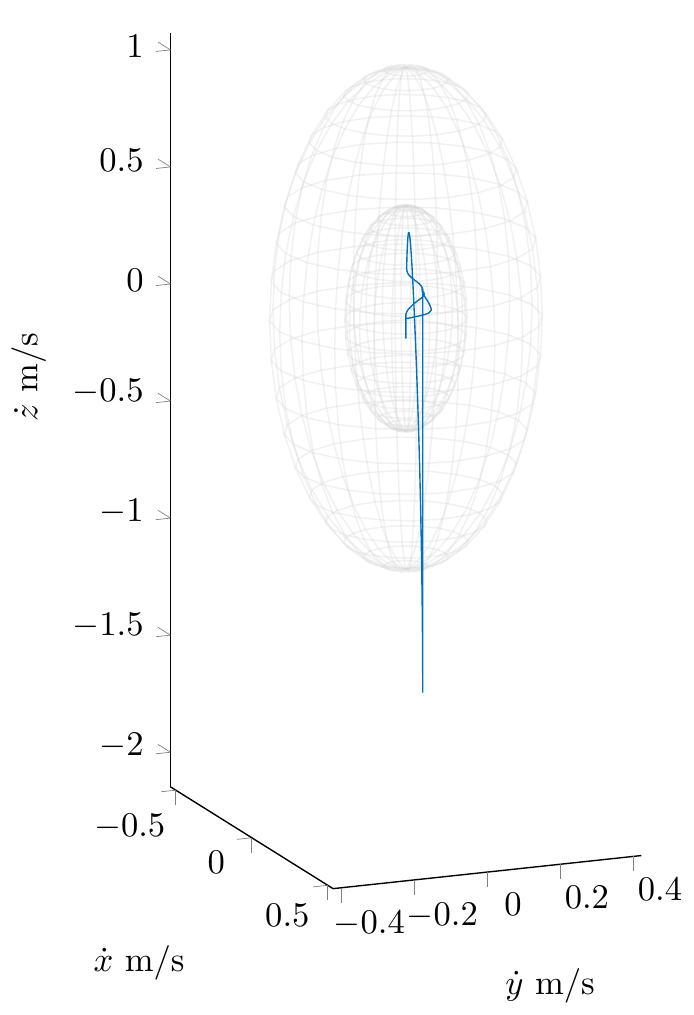}
	\end{subfigure}
	\caption{Projections of $\mathcal{E}_{SC}$(outer) and  $\mathcal{E}_{TC}$ (inner) through the several updated equilibrium points on the position and linear velocity spaces. The blue line represents the respective system response.} 
	\label{fig:3D_sim}
\end{figure*}

Figure \eqref{fig:3D_sim} illustrates the position and the linear velocities of the UAV respects to the projection of $\mathcal{E}_{SC}(x_{eq})$ (outer) and  $\mathcal{E}_{TC}(x_{eq})$ (inner) for several equilibrium points. The projection of $x_{eq}$ is always 0.

\begin{figure*}[!h]
	\centering
	\begin{subfigure}[]{} 
		\includegraphics[height=8cm, width=8cm]{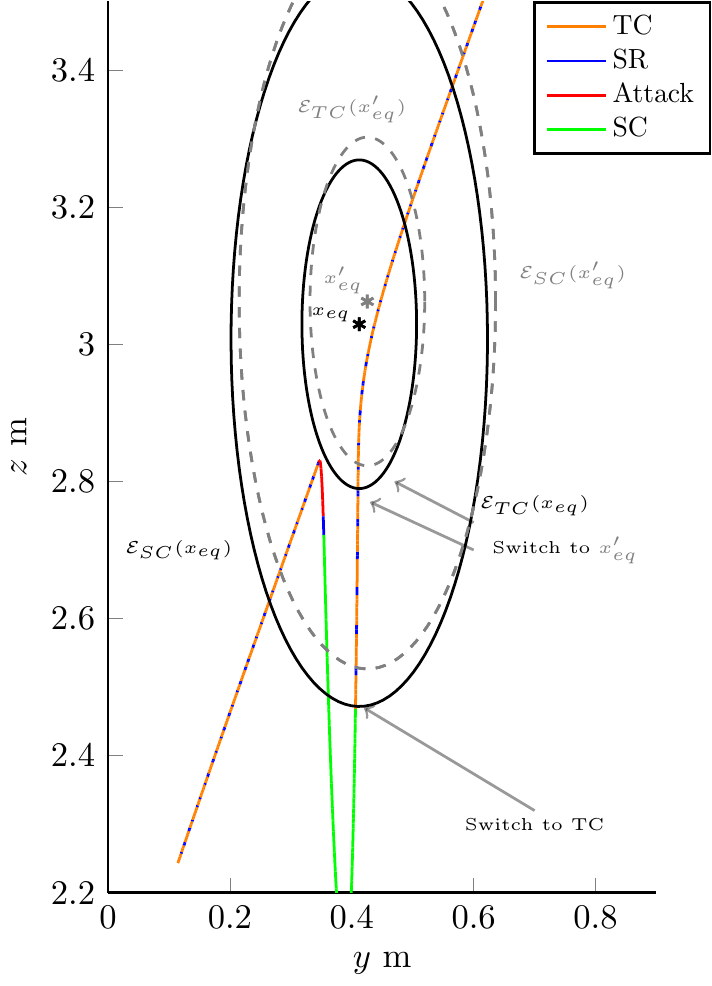}
	\end{subfigure}
	\hspace{1em} 
	\begin{subfigure}[]{} 
		\includegraphics[height=8cm,width=6cm]{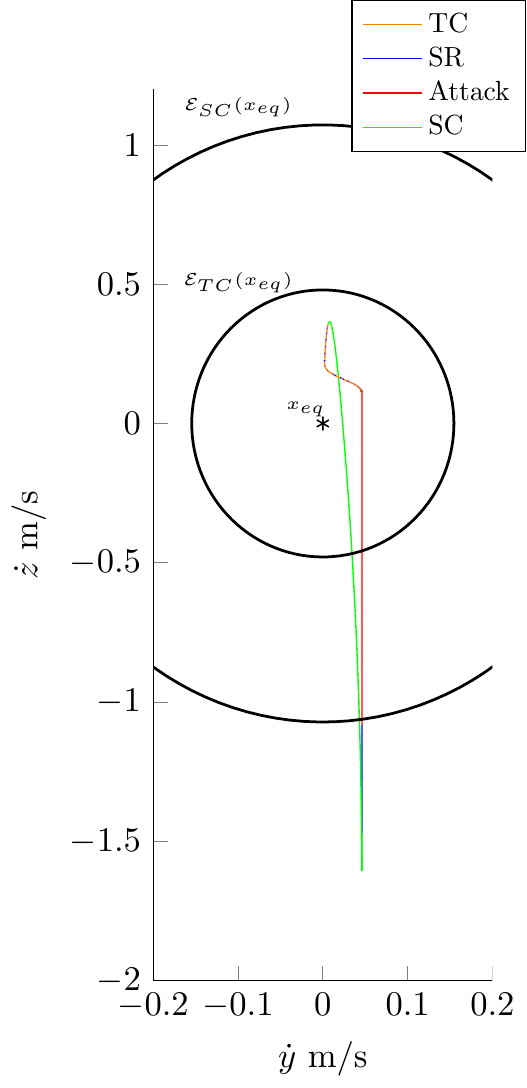}
	\end{subfigure}
	\caption{The switching modes behaviour in correspondence of the attack projected on the $y-z$ and $\dot y- \dot z$ planes. On the left it is illustrated the switch between two consecutive equilibrium points.} 
	\label{fig:2D-sim}
\end{figure*}

The projection of the sets $\mathcal{E}_{SC}(x_{eq})$ and  $\mathcal{E}_{TC}(x_{eq})$  in correspondence of the attach (Fig. \ref{fig:3D_sim}-left, blue ellipsoid) on the $y-z$ and plane $\dot y- \dot z$ is illustrated in Fig. \ref{fig:2D-sim} . That figure shows the behavior of the quadrotor during the several control modes. After the attack (red) and SR (blue), SC is active because the velocity (right) along $z$ hits $\mathcal{E}_{SC}(x_{eq})$ while the position is still inside that set (left). Instead, when the position goes back to $\mathcal{E}_{SC}(x_{eq})$, TC is active again. When the position hits $\mathcal{E}_{SC}(x_{eq})$, after SR the equilibrium point switches to $x_{eq}'$.  

\section{DISCUSSION}\label{sec:discussion}
This paper presents a general method for implementing software rejuvenation for secure tracking control based on an event-based reference point update scheme for the tracking controller and an invariant-based safety controller. An architecture for the onboard processor is presented that provides the necessary secure computation components to manage the software rejuvenation sequence, authentication and update of the reference points for the tracking controller and the mapping of the controller outputs to the system actuators.  This extends software rejuvenation to a broader class of CPS applications.

There are several directions for future research. There are several design trade-offs to be investigated, including the trade-offs among the control constraints and sizes of the reference point neighborhoods used for the tracking and safety control algorithms, the time allotted to tracking control and the control limits used during tracking control.  Disturbances and modeling uncertainties also need to be incorporated into the reachability algorithm used to determine the time bound on the uncertain control period. The integration of software rejuvenation with algorithms for attack detection and mitigation should also be investigated to extend the scope of security and robustness that can be achieved. Finally, we are currently moving the algorithm from simulation to a trusted real-time hypervisor platform for drone control and plan to report these results in future papers.

\section*{Acknowledgments}
\noindent 
Copyright 2018 IEEE. All Rights Reserved. \\
\noindent
This material is based upon work funded and supported by the Department of Defense under Contract No. FA8702-15-D-0002 with Carnegie Mellon University for the operation of the Software Engineering Institute, a federally funded research and development center. \\
\noindent
Carnegie Mellon® is registered in the U.S. Patent and Trademark Office by Carnegie Mellon University. \\
\noindent
DM18-1192

\bibliographystyle{unsrt}
\bibliography{references.bib} 

\appendix
\subsection{LQR Control}
The state space representation of the dynamics of a continuous-time LTI system is
\begin{equation}
\dot{x}(t)=A x(t)+Bu(t), \label{sys}    
\end{equation}
where $x \in \mathbb{R}^n$ is the state, and $u \in \mathbb{R}^p$ is the input.\\
A linear control that stabilizes \eqref{sys} is
\begin{equation}
    u(t)=-Kx(t) \label{contr}
\end{equation}
where $K$ is computed minimizing the cost function
\begin{equation}
J=\int_0^\infty x^T(\tau)Qx(\tau)+u^T(\tau)Ru(\tau)d\tau.   
\end{equation}
where $Q\geq 0$ and $R > 0$ are symmetric matrices. The minimum is achieved if
\begin{equation}
    K=R^{-1}B^TS \label{LQR}
\end{equation}
where $S$ is the solution of the associated Riccati algebraic equation \cite{lewis2012optimal}. Controller \eqref{contr} is a linear quadratic regulator. 

\subsection{Safe Control}
The associated closed-loop system is
\begin{equation}
    \dot{x}(t)=A_{SC}(x(t)-x_{eq}) \label{cloop}
\end{equation}
where $A_{SC}\triangleq A-BK$.

Given a convex polyhedral region $\mathcal{C} \subset \mathbb{R}^n$ of the state space given by
\begin{equation}
\mathcal{C}(x_{eq})=\left\lbrace x\; |\;  \xi_j^T (x-x_{eq})\leq 1,\; j=1,...,n_c \right\rbrace \label{C1}
\end{equation}
the largest positively invariant ellipsoid with respect to the equilibrium point $x_{eq}$ and associated to \eqref{cloop}
\begin{equation}
\mathcal{E}_{{\mathcal{C}}}(x_{eq})=\left\lbrace x \;|\;\Vert(x-x_{eq})\Vert_P^2  \leq 1 \right\rbrace \label{E_PSC}
\end{equation}
contained in $\mathcal{C}$  can be found solving the following maximization problem  \cite{boyd1994linear}:
\begin{equation}\label{safeinv}
\left\lbrace\begin{array}{l}
\mathrm{max\ log\ det} \bar Q\\
s.t.\ \bar Q A^{T}_{SC} + A_{SC} \bar Q\leq 0\\
\qquad\!\! \xi_j^T \bar Q \xi_j\leq 1,\; j=1,...,n_c\\
\qquad\!\! \bar Q>0
\end{array} \right.
\end{equation}
where $P=\bar Q^{-1}$ \cite{boyd1994linear}. Set \eqref{E_PSC} is a positively invariant set, since 
\begin{equation}
    \Vert(x-x_{eq})\Vert_P^2=(x-x_{eq})^T P (x-x_{eq}) 
\end{equation}
is a quadratic Lyapunov function for \eqref{cloop}. The set
\begin{equation}
\mathcal{E}_{SC}(x_{eq})=\left\lbrace x \;|\;\Vert(x-x_{eq})\Vert_P^2  \leq \epsilon_{SC} \right\rbrace \label{E_esc}
\end{equation}
with $0\leq\epsilon_{SC}\leq 1$ is a Lyapunov level-set where $\mathcal{E}_{SC}(x_{eq})\subseteq \mathcal{E}_{\mathcal{C}}(x_{eq})$.
From \eqref{C1} a convex polyhedron $\mathcal{E}_{{SC}}(x_{eq}) \subseteq \mathcal{C}_{{SC}}(x_{eq})$ is given by
\begin{equation}
\mathcal{C}_{SC}(x_{eq})=\left\lbrace x\; |\;  \xi_j^T (x-x_{eq})\leq \sqrt{\epsilon_{SC}},\; j=1,...,n_c \right\rbrace. \label{C2}
\end{equation}

\subsection{Computation of $T_{UC}$}
Given $\epsilon_{SC}$, the computation of $T_{UC}$ consists in checking if the reachable set
\begin{eqnarray}
    &\hspace{-0.7cm} \mathcal{R}(\mathcal{C}_{SC}(x_{eq}),t;\mathcal{U})=  \{x| e^{At}x_0 + \int_0^t e^{A(t-\tau)}Bu(\tau)d\tau,  \\
 &\hspace{2cm}x_0 \in \mathcal{C}_{SC}(x_{eq}), \ u(\tau) \in \mathcal{U}, \ 0 \leq \tau \leq t \}. \label{reach_set}  
\end{eqnarray}
is contained in $\mathcal{E}_{\mathcal{C}}(x_{eq})$. The reach set \eqref{reach_set} is computed considering any value of input in $\mathcal{U}$ given to the system $\eqref{sys}$, starting from any point in $\mathcal{C}_{SC}(x_{eq})$.

An approximation of \eqref{reach_set} is obtained considering a method based on the computation of the supporting hyperplans for polytopic sets \cite{hwang2005polytopic}. The supporting hyperplanes to $\mathcal{R}(\mathcal{C}_{SC}(x_{eq})$ are
\begin{equation}
    v^+_j(x,t)=\xi_j(t)x,\qquad j=1,...,n_c \label{v+}
\end{equation}
where $\xi_j(0)= \xi_j$. Using the maximum principle of the optimal control \cite{varaiya2000reach}, the normal direction of $\xi_i$ is
\begin{equation}
    \xi_j(t)=e^{-A^T\tau}\xi_j. \label{normal_dir}
\end{equation}
Representing $\mathcal{U}$ as convex combination of $n_u$ vertexes $u_1,...,u_{n_u}$, the over-approximation of $\mathcal{R}(\mathcal{C}_{SC}(x_{eq}),t,\mathcal{U})$ is \cite{kurzhanski2001dynamic}
\begin{eqnarray}
    &\hspace{-1cm}\mathcal{R}^+(\mathcal{C}_{SC}(x_{eq}),t,\mathcal{U})= \bigcap_{j=1}^{n_c}\left\lbrace x|v^+_j(x, t) \leq \right. \nonumber \\
    &\hspace{2cm}\left. \int_0^{ t} \max_i\;\langle\xi_j(\tau),Bu_i \rangle d\tau + \right.\nonumber \\
    &\hspace{3cm}\left. \max_{x(0)\in \mathcal{\mathcal{C}_{SC}(x_{eq})}} v^+_j(x(0),0) \right\rbrace. \label{V+1} 
\end{eqnarray}
The convex polyhedral set \eqref{V+1} can be also represented as polytope, 
\begin{eqnarray}
    &&\hspace*{-1cm}\mathcal{R}^+(\mathcal{C}_{SC}(x_{eq}),t,\mathcal{U})=\nonumber\\
    &&\hspace*{-1cm}\left\lbrace x \in \mathbb{R}^n\;|\; x=\sum_{k=1}^{n_v}\alpha_h x_k,\; \sum_{k=1}^{n_v}\alpha_k =1\; \alpha_k\geq 0 \right\rbrace
\end{eqnarray}
then to check if for a given $T_{UC}$, $\mathcal{R}^+(\mathcal{C}_{SC}(x_{eq}),t,\mathcal{U}) \subseteq \mathcal{E}_{\mathcal{C}}(x_{eq})$ it is sufficient to make that check only in correspondence of the verteces  $x_k$.

\end{document}